\documentclass[12pt]{article}
\usepackage{epsfig}

\newcommand{\la}{\langle}
\newcommand{\ra}{\rangle}
\begin{document}


\begin{flushright}
\bf IFJPAN-V-05-04\\
\bf CERN-PH-TH/2005-094
\end{flushright}

\vspace{1mm}
\begin{center}
{\Large {\bf
  mFOAM-1.02: A Compact Version\\ 
  of the Cellular Event Generator FOAM$^{\star}$}
}
\end{center}

\vspace{1mm}

\begin{center}
{\bf S.~Jadach}\\
\vspace{1mm}
{\em Institute of Nuclear Physics, Academy of Sciences,\\
  ul. Radzikowskiego 152, 31-342 Cracow, Poland,}\\
{\em and}\\
{\em CERN Department of Physics, Theory Division\\
CH-1211 Geneva 23, Switzerland}\\
\vspace{1mm}
{\rm and}\\
{\bf P.~Sawicki} \\
\vspace{1mm}
{\em Institute of Nuclear Physics, Academy of Sciences,\\
  ul. Radzikowskiego 152, 31-342 Cracow, Poland}\\
\end{center}

\vspace{2mm}
\begin{abstract}
The general-purpose self-adapting Monte Carlo (MC) event generator/simulator
{\tt mFOAM} (standing for mini-FOAM) 
is a new compact version of the {\tt FOAM} program, 
with a slightly limited functionality with respect to its parent version.
On the other hand, {\tt mFOAM} is easier to use for the average user.
This new version is fully integrated with the ROOT package,
the C++ utility library used widely in the particle physics community. 
The internal structure of the code is simplified 
and the very valuable feature of
the persistency of the objects of the {\tt mFOAM} class is improved.
With the persistency at hand, it is possible to record very easily 
the complete state of a MC simulator object
based on {\tt mFOAM} and ROOT into a disk-file at any stage of its
use: just after object allocation,
after full initialization (exploration of the distribution), or
at any time during the generation of the long series of MC events.
Later on the MC simulator object can be easily restored
from the disk-file in the ``ready to go'' state.
Objects of {\tt TFoam} class can be used as a stand-alone solution to many
everyday problems in the area of the Monte Carlo simulation,
or as building blocks in large-scale MC projects, 
taking full advantage of the object-oriented technology and persistency.
\end{abstract}

\vspace{2mm}
\begin{center}
\em To be submitted to Computer Physics Communications
\end{center}

\vspace{5mm}

\vspace{1mm}
\begin{flushleft}
{\bf IFJPAN-V-05-04\\
 \bf CERN-PH-TH/2005-094\\
     June~2005}
\end{flushleft}

\vspace{2mm}
\footnoterule
\noindent
{\footnotesize
$^{\star}$Supported in part by EU grant MTKD-CT-2004-510126,
  in partnership with CERN PH/TH Division.
}

\newpage
\noindent{\bf PROGRAM SUMMARY}
\vspace{10pt}

\noindent{\sl Title of the program:}\\
{\tt mFOAM (mini FOAM)}, version 1.02.

\noindent{\sl Computer:}\\
Most Unix workstations, supercomputers and PC.

\noindent{\sl Operating system:}\\
Most UNIX systems, Linux and Windows. \\ 
Application programs were thoroughly tested under Red Hat Linux 7.x, 
CERN Scientific Linux 3.02, Fedora Linux FC3, UNIX IRIX-6.5.\\
At present {\tt mFOAM} is distributed with the ROOT package
(version 4.04 and later).

\noindent{\sl Programming languages used:}\\
ANSI C++.

\noindent{\sl High-speed storage required:}\\
Depends on the complexity of the problem.
For the default 2000 cells it is about 25 MB
while for 100,000 cells it allocates about 35 MB.
These data are for running from CINT command
line and include also memory consumption by CINT itself.

\noindent{\sl No. of lines in combined program and test deck:}\\
{\tt mFOAM}-1.02 2776 lines of C++ code.

\noindent{\sl Nature of the physical problem:}\\
Monte Carlo integration or generation of unweighted (weight equals 1)
events with a given probability distribution is a standard problem 
in many areas of research, ranging from high-energy physics to 
economy. In any library of general utilities it is highly desirable to include 
a general-purpose numerical tool (program) with the MC 
generation algorithm featuring the built-in capability of 
automatically adjusting generation procedure to an arbitrary pattern 
of singularities in the generated distribution. 
Our primary goal is the simulation of the differential distribution in the
multiparticle Lorentz-invariant phase space for the purpose of comparison
between Quantum Field Theory prediction, and experiments in the high-energy 
experiments.
However, the solution may have a much wider area of applications.

\noindent{\sl Method of solution:}\\
In the algorithm, 
a grid of cells, called ``foam'', is built in the process of the binary
split of the cells. The resulting foam is adapted automatically to the shape 
of the integrand in such a way that 
the resulting ratio of the average weight to
maximum weight or the variance to average weight is minimized.

\noindent{\sl Restrictions on the complexity of the problem:}\\
Consumption of computer resources depends on the complexity of the problem.
The use of the program is limited to about a million of cells 
for a relatively small 
number of dimensions ($\leq 20$)
in view of the memory and CPU time restrictions 
of a modern desktop computer.

\noindent{\sl Typical running time:}\\
The CPU time necessary to build up a foam of cells depends strongly 
on the number of dimensions and the requested number of cells. On
the PC with a 1.6~GHz Intel processor, it takes about 10 seconds to build
a hyperrectangular grid of 10,000 cells for simple 3-dimensional 
distribution.

\newpage

\section{Introduction}

The present program {\tt mFOAM},
and the {\tt FOAM} program of Refs.~\cite{Jadach:2002kn,Jadach:1999sf},
from which {\tt mFOAM} is derived,
are both examples of a general-purpose self-adapting Monte Carlo
simulator/integrator.
Let us briefly recapitulate main features of {\tt FOAM},
which are shared with the present project.
In the cellular algorithm of {\tt FOAM}, points are generated randomly in 
the multi dimensional space according to an arbitrary, user-defined,
unnormalized probability distribution function (PDF) $\rho(x)$.
The algorithm works in two stages:
{\em exploration } and {\em generation}. In the exploration stage
the shape of the distribution function is explored using MC methods,
dividing the integration domain into a system of cells referred to as ``foam''.
The foam of cells is produced in a recursive process of binary splittings of 
the cells starting from the root cell, which can be a single $k$-dim
hyperrectangle, an $n$-dim simplex or a Cartesian product of both. 
In {\tt mFOAM} we restrict ourselves to hyperrectangles.
The PDF $\rho(x)$ is approximated by another PDF
$\rho '(x)$, which is equal to a constant within each cell. 
The main aim of the process of the foam evolution through binary splittings is 
to minimize either the ratio of the variance of the
weight distribution to the average weight  $\sigma/\la w \ra$,
or the ratio of the maximum weight to the average
weight $w_{\max}/ \la w \ra$,
where  $w=\rho (x) / \rho^{'} (x)$ is the Monte Carlo weight.

In the generation stage every single MC weighted event
is generated as follows: first a cell is chosen randomly 
and next, within this cell, a point (MC event) is generated
according to an uniform distribution equal to $\rho '$
and finally the MC weight 
$w=\rho(x)/\rho'(x)$ is evaluated.
As usual, the rejection method may turn these weighted events into 
weight-one events, with a certain rejection rate (inefficiency).
The main aim of the rather sophisticated
cell-splitting algorithm of {\tt FOAM} (exploration phase) is
the reduction of $w_{\max}/ \la w \ra$, assuring a low rejection rate.
Another option is the variance-reduction providing for self-adapting
MC method of precise evaluation of the integrals.
In either case, the value of the integrand is already known approximately
from the exploration stage and can be 
estimated with even better precision in the generation phase. 

It is instructive to compare the cellular algorithm of {\tt FOAM} 
to algorithms used by two older programs 
in the family of self-adapting MC tools: 
VEGAS \cite{Lepage:1978sw} and MISER \cite{Press:1989vk}.
VEGAS primarily implements the so-called importance sampling
(variance-reducing) method.
It approximates the exact distribution by a multidimensional
sampling function $g$. The function $g$ is separable by construction,
i.e. $g(x_1,x_2, \ldots, x_n)= g_1(x_1) g_2(x_2),\ldots,g_n(x_n)$.
Owing to this feature, the function $g$ can be stored  effectively
in the computer memory  as a collection of $n$ (one for each dimension)
histograms with $K$ bins, without an explosion in the total number of 
bins, which would in general grow like $K^n$.
The sampling distribution is constructed iteratively, step by step,
by means of making a number of Monte Carlo
explorations over the integration region,
while inspecting $n$ 1-dimensional histograms of the
projections the distribution function, each for one dimension. 
These histograms are used to define the new improved function $g_i$,
which in turn are used to generate MC points in the next iteration.
In principle, after a few iterations, one obtains 
the reference distribution $g$ approximating the PDF. 
An estimated of the value of the integral over PDF is also obtained.
In practice the performance of VEGAS depends heavily on the goodness of
the factorizability assumption for a given PDF. 
Generally, VEGAS turns out to be quite efficient
for many distributions (integrands) featuring a single well localized peak.

The MISER program%
\footnote{Unfortunately, the MISER algorithm was overlooked in the previous 
  papers on the {\tt FOAM} project.}
is based on the idea of the ``recursive stratified sampling'' and employs 
the technique of variance reduction similar to that in {\tt FOAM}.
It explores the PDF until a fixed maximal number of available 
function evaluations $N$ is exhausted.
In the very beginning $N$ is allocated to the root cell being a hypercube
and later on redistributed among the daughter cells.
In the simplest variant the starting hypercube is divided by bisecting 
it across one of the edges into two sub-cells of equal volume%
\footnote{A quite similar 2-dimensional algorithm
  is also present in the MC program LESKO, of ref.~\cite{Jadach:1991ty},
  and in other programs; see ref.~\cite{Jadach:1999sf} for more references.}.
The division plane is chosen
by examining all possible $n$ bisections of the $n$-dimensional cell
and selecting the one that minimizes the resulting total
variance of the two cells.
Similarly as  in {\tt FOAM},
the variances  are estimated cell by cell
during a short MC survey with a small fraction of  ``allocated'' 
events for this cell. 
The remaining pool of unexploited function calls is allocated 
to the resulting sub-cells in a proportion that fulfills 
the condition for minimum variance. The whole procedure is 
repeated for each of the two sub-cells and continues recursively 
until the number of ``allocated function calls'' in a given cell falls 
below some predefined limit. In each cell the estimation 
of the integral  is obtained by means of the plain MC method.
At the end, the results 
for all cells are combined together to obtain the final value of 
the integrand and the error estimate.

{\tt FOAM} employs a combination of both techniques: importance and
stratified sampling.
Contrary to VEGAS, there is no assumption in the {\tt FOAM} 
algorithm about the factorizability of the distribution (integrand).
In the variance reduction mode {\tt FOAM} resembles MISER,
but it employs a different,
far more sophisticated cell division algorithm;
the division plane of the cell is not at the half-point of the edge, but
is optimized.
The algorithm of {\tt FOAM} has
passed many practical tests and proved its efficiency in several
problems in high-energy physics;
see for instance~\cite{Placzek:2003zg,Jadach:2005bf}.
The foundations of the {\tt FOAM} algorithm are well consolidated and
our current work concentrates mainly on the updates of earlier 
implementations and improvements of the efficiency and functionality.
For a detailed description of the algorithm of {\tt FOAM}
version 2.05 we refer the
interested reader to Refs.~\cite{Jadach:2002kn} and~\cite{Jadach:1999sf}.

The use of the original {\tt FOAM} program \cite{Jadach:1999sf}
has been mainly limited by the memory consumption. 
{\tt FOAM}~v.2.05 divides the $n$-dimensional
parameter space into hyperrectangular or simplical cells. 
Final MC efficiency increases mainly with the requested maximum number of 
cells $N_c$, so it is very important to economize
on the memory used by single cell in order reach a higher number of cells.
For the hyperrectangular grid of cells a
memory saving arrangement algorithm of coding cells in the memory 
was found~\cite{Jadach:2002kn}.
It reduces memory consumption down to a mere
80 bytes/cell, independently of space dimension $n$.
The present version, limited to hyperrectangles,
profits from this memory-saving algorithm of recording the cell parameters.
We would like to mention, that in the mean-time a similar memory-saving
algorithm has been also found and implemented for simplices.
It will be included in the forthcoming version 
2.06 of the {\tt FOAM}~\cite{foam2-6}.

The unspoken assumption in {\tt mFOAM} is that the calculation of the PDF
is cheap in terms of CPU time. 
This is often true in practice.
If not, then {\tt mFOAM} may be used to model the main features
of the singularities in the PDF and the fine details, 
which can be CPU-costly,
are then added by extra MC weight during the MC run, after the exploration.
However, in order to deal better with the cases of PDFs which are costly
in terms of CPU and feature relatively mild peaks, one should introduce
in the future development of {\tt mFOAM} the possibility
to limit the total number of PDF calls,
in addition to limiting the number of cells.

The paper is organized as follows: Section 2 describes
changes in basic classes  and their functionality.
Section 3 describes the configuration of {\tt mFOAM}.
Section 4 discusses the usage of {\tt mFOAM} classes 
under the ROOT system. Conclusions follow. 

\section{Description of {\tt mFOAM} code} 

{\tt mFOAM} (mini FOAM) is a new version of {\tt FOAM} with slightly
limited functionality, well integrated with ROOT~\cite{root:1997}.
Our principal aim is to provide a compact and easy to use tool,
for numerical Monte Carlo generation and integration of PDFs with arbitrarily
complicated structure of peaks,
in the number dimensions limited up to say 20.
With the increasing popularity of ROOT in high-energy
community we believe that this implementation tied up with ROOT will
attract the interest of the new users who already exploit ROOT
in their daily work.

Let us comment on our decision of removing the simplical cells from 
the {\tt mFOAM} algorithm and the code.
It was done because of an empirical observation
(based on practical experience with the wide range of the distributions)
that the use of simplical cells was usually giving rise to worse MC efficiency
than that of hyperrectangular cells.
In addition, maintaining simplical cells increases 
complexity of the source code.

The main motivation for the closer integration of {\tt mFOAM}
with the ROOT system was to profit fully from the {\em persistency mechanism}
for its objects and help users who already use ROOT daily.
Also, thanks to the closer integration with ROOT, the code of {\tt mFOAM} gets
more compact, since the internal histogramming
and other low-level structures are replaced
by the well tested ROOT facilities. 
Altogether, we have managed
to reduce significantly the total size of code (by about 50\%)
and its complexity as well,  with respect to the original {\tt FOAM},
at the same time improving its stability.

Obviously, the above improvements and gains are purely technical,
nevertheless they are very important, if object of the {\tt mFOAM} class
are to be used as ``rock solid'' building blocks in any more complex,
large scale, Monte Carlo projects.


\begin{table}[!bt]
\centering
\begin{small}
\begin{tabular}{|l|p{12.0cm}|}
\hline 
Class & Short description \\
\hline \hline
{\tt TFoamIntegrand} & Abstract class for the integrand function \\
{\tt TFoamVect } & Utility class of vectors 
                   with dynamic allocation of memory \\
{\tt TFoamCell } & Class representing the single-cell object \\
{\tt TFoam  } & Main class of mFOAM. The entire MC generator\\
{\tt TFoamMaxwt }& Monitors MC weight, measures performance of the MC run \\
\hline 
\end{tabular}
\end{small}
\caption{\sf Summary on C++ classes of {\tt mFOAM}.}
  \label{tab:classes}
\end{table}

{\tt mFOAM}, like its ancestor, is written fully in the 
object-oriented programming (OOP) style in the C++
programming language. The classes of the {\tt mFOAM} program are listed in 
Table~\ref{tab:classes}. Some classes present in {\tt FOAM-2.05} have been 
removed, because they are needed only for the simplical cells.
The remaining classes
changed their names to comply with the ROOT naming conventions.
For the same reason,
names of preserved data members now begin with the letter ``f''.   
Two basic classes, {\tt TFoam} and {\tt TFoamCell}, are greatly simplified 
by the removing all of simplical structure. 
All other remaining classes have the same functionality
as in  {\tt FOAM} version 2.05. 
In particular, an abstract base class {\tt TFoamIntegrand} 
provides the user interface to any user-provided PDF.
Classes {\tt TFoamVect} and {\tt TFoamMaxwt} are unmodified 
auxiliary  utility classes. 
In {\tt mFOAM} we use the library of random 
generators of ROOT; the {\tt TPSEMAR} class of {\tt FOAM} is removed.
All classes of {\tt mFOAM}
inherit I/O capabilities  from ROOT's {\tt TObject} class.

As already advertised, we have payed special attention to the persistency issue.
Generally, it is not trivial to get full persistency for the {\tt mFOAM} 
and {\tt FOAM} classes, mainly because of the intensive use of the pointers
in the coding of the linked binary trees of the foam cells.
All these problems are now solved efficiently with the help of 
the ROOT pointer classes.
Consequently, any object of the {\tt mFOAM} class can be written more easily at 
any time into disk and restored later on, with the help
of the ``automatic streamers'' generated by ROOT.
In this way, generation of the MC events can be easily stopped and resumed.
When the MC generation of the series of events is resumed,
then MC generation continues
as if there was no disk-read and disk-write in the meantime.

A simple persistent abstract class (interface)
representing any user-defined PDF is available. 
We refer the reader to Section~\ref{sec:usage} for 
a number of explicit examples/templates how to exploit it.

Let us now characterize briefly the role of most important classes
in the implementation of the {\tt mFOAM} algorithm.

\subsection{{\tt TFoam} class}


\begin{table}[!ht]
\centering
\begin{small}
\begin{tabular}{|l|p{11.0cm}|}
\hline
{\tt TFoam} member & Short description  \\ 
\hline\hline
  TString  fVersion$^g$ & Actual version of the {\tt mFOAM} (like 1.02m)\\
  TString  fDate        & Release date of the {\tt mFOAM}\\
  TString  fName        & Name of a given instance of the {\tt TFoam} class\\
  Int\_t fDim$^{s,g}$      & Dimension of the integration space\\
  Int\_t fNCells$^s$    & Maximum number of cells\\
  Int\_t fRNmax         & Maximum number of random numbers generated at once\\
\hline
  Int\_t fOptDrive$^s$  & Optimization =1,2 for variance or maximum weight 
                                           reduction\\
  Int\_t fChat$^s$      & =0,1,2 chat level in output; =1 for normal output\\
  Int\_t fOptRej$^s$    & =0 for weighted events; =1 for unweighted events in MC 
generation\\
\hline
  Int\_t  fNBin$^s$     & No. of bins in edge histogram for cell MC exploration\\
  Int\_t  fNSampl$^s$   & No. of MC events, when dividing (exploring) cell\\
  Int\_t  fEvPerBin$^s$ & Maximum number of effective ($w=1$) events per bin\\
  Double\_t fMaxWtRej$^s$;    &Maximum weight in rejection for getting $w=1$ events\\  
\hline
\end{tabular}
\end{small}
\caption{\sf Data members of the {\tt TFoam} class.
  Associated setters and getters marked as superscripts $s$ and $g$.}
  \label{tab:TmFOAMmembers1}
\end{table}

\begin{table}[!ht]
\centering
\begin{small}
\begin{tabular}{|l|p{90mm}|}
\hline
{\tt TFoam} member & Short description  \\ 
\hline
\multicolumn{ 2}{|c|}{{ Provision for the multibranching } }\\
\hline
  Int\_t    *fMaskDiv     &![fDim] Dynamic mask for cell division\\
  Int\_t    *fInhiDiv     &![fDim]  Flags inhibiting cell division \\
  Int\_t     fOptPRD      &Option switch for predefined division, for quick check\\
  TFoamVect **fXdivPRD    &!Lists of division values encoded in one vector per direction\\
\hline
\multicolumn{ 2}{|c|}{{ Geometry of cells } }\\
\hline
  Int\_t fNoAct           &Number of active cells\\
  Int\_t fLastCe          &Index of the last cell\\
  TFoamCell **fCells      &[fNCells] Array of ALL cells\\
\hline
\multicolumn{ 2}{|c|}{{ Monte Carlo generation } }\\
\hline
  TFoamMaxwt *fMCMonit;   &Monitor of the MC weight for measuring MC efficiency\\  
  TRefArray *fCellsAct &Array of pointers to active cells. \\
  Double\_t *fPrimAcu     &[fNoAct] Array of cumulative $\sum_{i=1}^k R'_i$ \\
  TObjArray *fHistEdg  &Histograms of $w$, one for each edge \\
  TObjArray *fHistDbg  &Histograms for debug  \\
  TH1D      *fHistWt;  &Histograms of MC weight \\
\hline
\multicolumn{ 2}{|c|}{{ Externals } }\\
\hline
  TMethodCall* fMethodCall$^s$ & !ROOT's pointer to global distribution function \\
  TFoamIntegrand *fRho$^{g,s}$ & Pointer to class with distribution function \\
  TRandom         *fPseRan$^{g,s}$  &Generator of the uniform pseudo-random numbers\\
\hline
\multicolumn{ 2}{|c|}{{ Statistics and MC results } }\\
\hline
  Long\_t fNCalls$^g$      &Number of function calls\\
  Long\_t fNEffev$^g$      &Total No. of effective $w=1$ events in build-up\\
  Double\_t fSumOve         &Sum of overweighted events \\
  Double\_t fSumWt, fSumWt2&Sum of weight $w$ and squares $w^2$\\
  Double\_t fNevGen        &No. of MC events\\
  Double\_t *fMCvect        &[fDim] MC vector \\
  Double\_t fMCwt           &MC weight \\
  Double\_t fWtMax, fWtMin &Maximum/Minimum weight (absolute)\\
  Double\_t fPrime$^g$     &Primary integral $R'$, ($R=R' \langle w \rangle$)\\
  Double\_t fMCresult      &True integral $R$ from the cell exploration MC\\
  Double\_t fMCerror       &and its error\\
  Double\_t *fRvec          &[fRNmax] random number vector \\
\hline
\multicolumn{ 2}{|c|}{{ Working space for cell exploration } }\\
\hline
  Double\_t *fAlpha       &[fDim] Internal parameters of the h-rectangle: $0<\alpha_i<1$\\
\hline
\end{tabular}
\end{small}
\caption{\sf Data members of the {\tt TFoam} class. Cont.}
  \label{tab:TmFOAMmembers2}
\end{table}

\begin{table}[hp]
\centering
\begin{small}
\begin{tabular}{|l|p{80mm}|}
\hline
{\tt TFoam} method & Short description  \\ 
\hline
\multicolumn{ 2}{|c|}{ Constructors and destructors }\\
\hline
  TFoam()                          & Default constructor (for ROOT streamer)\\
  TFoam(const Char\_t *)               & User constructor\\
  $\tilde{\mbox{}}$ TFoam()           & Explicit destructor\\
  TFoam(const TFoam\&)             & Copy Constructor  NOT USED\\
  TFoam\& operator=(const TFoam\& )& Substitution      NOT USED \\
\hline
\multicolumn{2}{|c|}{ Initialization, foam build-up }\\
\hline
  void Initialize()               & Initialization, allocation of memory\\
  void SetRho(TFoamIntegrand *)   & Sets the pointer to distribution function\\
  void ResetRho(TFoamIntegrand *)  & Resets the pointer to distribution function\\
  void SetRhoInt(void *)           & Sets the pointer to user-defined global function \\
  void SetPseRan(TRandom*)         & Sets the pointer to r.n.g. \\
  void ResetPseRan(TRandom*)       & Resets the pointer to r.n.g.  \\
  void InitCells(void)             & Initializes memory for cells and starts exploration\\
  void Grow(void)                  & Adds new cells to foam, until buffer is full\\
  Int\_t  Divide(TFoamCell *)      & Divides cell into two daughters\\
  void Explore(TFoamCell *Cell)    & MC exploration of cell main subprogram\\
  void Carver(Int\_t\&,Double\_t\&,Double\_t\&)& Determines the best edge,
                                     $w_{\max}$ reduction\\
  void Varedu (Double\_t[~],       &   \\
  Int\_t\&,Double\_t\&,Double\_t\&)& Determines the best edge, $\sigma$ reduction\\
  Long\_t PeekMax(void)               & Chooses one active cell, used in {\tt Grow}\\
  void MakeAlpha(void)             & Generates rand. point inside h-rectangle\\
  Int\_t  CellFill(Int\_t, TFoamCell*) & Fills next cell and returns its index\\
  void MakeActiveList(void)        & Creates table of all active cells\\
  void SetInhiDiv(Int\_t, Int\_t )       & Sets inhibition of cell division along certain edge \\
  void SetXdivPRD(Int\_t, Int\_t, Double\_t[]); & Sets predefined division points\\
  Double\_t Eval(Double\_t *)            & Evaluates value of the distribution function \\ 
\hline
\multicolumn{ 2}{|c|}{ Generation }\\
\hline
  void   MakeEvent(void)           & Makes (generates) single MC event\\
  void   GetMCvect(Double\_t *)       & Provides generated random MC vector\\
  Double\_t GetMCwt(void)             & Provides MC weight\\
  Double\_t MCgenerate(Double\_t *MCvect)& All the above in single method\\
  void GenerCel2(TFoamCell *\&)       & Chooses one cell with probability $\sim R'_j$\\
\hline
\multicolumn{ 2}{|c|}{ Finalization, reinitialization }\\
\hline
  void Finalize(Double\_t\&, Double\_t\&)  & Prints summary of MC integration\\
  void GetIntegMC(Double\_t\&, Double\_t\&)& Provides MC integral\\
  void GetIntNorm(Double\_t\&, Double\_t\&)& Provides normalization\\
  void GetWtParams(const Double\_t, & \\
  ~~~~Double\_t\&, Double\_t\&, Double\_t\&)  & Provides MC weight parameters\\
\hline
\multicolumn{ 2}{|c|}{ Debug }\\
\hline
  void CheckAll(const Int\_t)   & Checks correctness of the data structure\\
  void PrintCells(void)      & Prints all cells\\
\hline
\end{tabular}
\end{small}
\caption{\sf Methods of the {\tt TFoam} class.}
  \label{tab:TmFOAMmethods1}
\end{table}

{\tt TFoam} is the main class. Each instance of the {\tt TFoam} class is 
a separate, independent MC generator.  In Tables~\ref{tab:TmFOAMmembers1} and 
\ref{tab:TmFOAMmembers2}, we provide a full list of data members of the
class {\tt TFoam} and their short description. Most of the methods 
(procedures) of the class {\tt TFOAM} are listed in
Table~\ref{tab:TmFOAMmethods1}. We omitted in this table ``setters'' 
and ``getters'', which provide access to some data members, and
simple inline functions, such as {\tt sqr} for squaring a
{\tt Double\_t} variable. Data members that are served by the
setters and getters are marked in Tables~\ref{tab:TmFOAMmembers1} and
\ref{tab:TmFOAMmembers2} by the superscripts ``$s$'' or/and ``$g$''.
We followed closely the ROOT naming conventions and decided to use appropriate 
ROOT types instead of raw C number types. 
In this way we assure the portability of our code to 
the forthcoming generation of inexpensive 64-bit processors.
    
Below we briefly describe the functionality of the most important methods
in the {\tt TFoam} class.

\subsubsection{Constructor}
The  {\tt TFoam(const Char\_t *)} constructor creates the  {\tt TFoam}
object whose name is given by its argument.
For example the following line of code creates an instance of {\tt mFOAM}
generator named {\tt FoamX}:
\begin{verbatim}
TFoam   *FoamX    = new TFoam("FoamX");   // Create Simulator
\end{verbatim}
The main role of the 
constructor is to initialize data members to their default values 
-- no memory allocation is done at this stage. The principal configuration 
parameters can be optionally changed by using setter methods 
(this is described in Sect.~\ref{configuring} ). 

\subsubsection{Setting distribution function and random number generator}
The user should also provide her/his own 
unintegrated non-negative probability distribution function (PDF).
Note that the PDF may be discontinuous.
{\tt mFOAM} can cope with integrable infinite singularities in the PDF.
However, we do not really recommend to use it for such cases.

Two methods were available for providing a PDF object to an {\tt mFOAM} object: 
{\tt SetRho(TFoamIntegrand *)} sets the  pointer of the PDF object
through the abstract class {\tt TFoamIntegrand} pointer (interface).
The user can also provide a global PDF, making it
available to the {\tt mFOAM} object by calling the method 
{\tt SetRhoInt(void *)}.
A detailed description of how to implement all kinds of PDFs is given
in Sect.~\ref{example}. 

The random number generator (RNG) object is created  by the user and 
set as a pointer in the {\tt SetPseRan (TRandom *)} method;
see explicit examples in  Sect.~\ref{sec:usage}.

How to organize the interrelation between 
the RNG and PDF objects of {\tt TRandom} and {\tt TFoamIntegrand} classes,
serving several objects of the {\tt mFOAM} class without destroying
the persistency, will be discussed in Sect.~\ref{sec:extern}.

\subsubsection{Initialization step methods}
To begin the process of the foam build-up, the user should invoke 
the {\tt Initialize()} method. 
The method  {\tt InitCells} initializes the memory storage for 
cells and begins the exploration process starting from the root cell. The empty 
cells are allocated/filled using  {\tt CellFill}. The procedure {\tt Grow} 
which loops over cells, picks up the cell with the biggest 
``driver integral'' (see Ref.~\cite{Jadach:2002kn} for explanations)
with the help of the {\tt PeekMax} procedure. The chosen cell is 
split using the {\tt Divide} procedure.

\begin{figure}[!ht]
\begin{center}
\epsfig{file=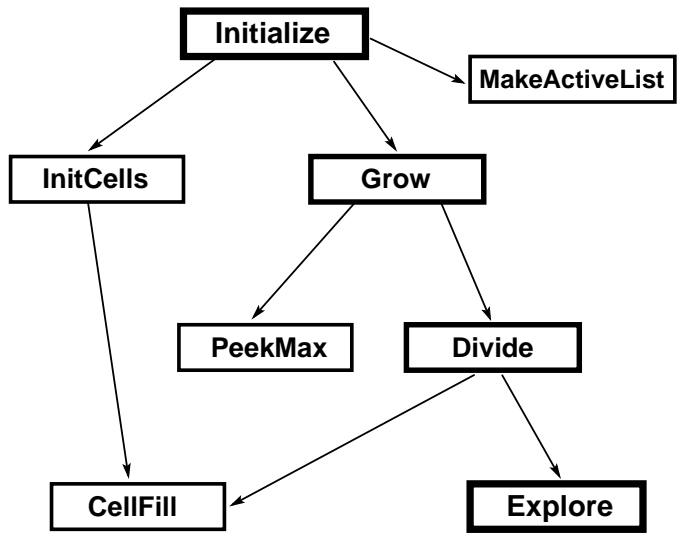,width=80mm,angle=270}
\end{center}
\caption{\sf
  Calling sequence of the {\tt mFOAM} procedures during 
  the foam build-up (initialization).
}
\label{fig:initialize}
\end{figure}

Subsequently, the procedure {\tt Explore} called by {\tt Divide}
(and by {\tt InitCells} for the root cell) does the most important 
job in the {\tt mFOAM} build-up: it performs a low statistics MC 
exploration run for each newly allocated daughter cell.
It calculates how profitable the future split of the cell will be
and defines the optimal cell division geometry with the help
of the {\tt Carver} or {\tt Varedu} procedures,
for maximum weight or variance optimization respectively.
All essential results of the exploration are written into 
the explored cell object. At the very end of the foam build-up,
{\tt MakeActiveList} is invoked to create a list of pointers to
all active cells, for the purpose of the quick access during 
the MC generation. The procedure {\tt Explore} uses  
{\tt MakeAlpha}, which provides random coordinates inside 
a given cell with the uniform distribution. The above sequence 
of the procedure calls is depicted in Fig.~\ref{fig:initialize}.

\subsubsection{MC event generation step methods}
The MC generation of a single MC event is done by invoking
{\tt MakeEvent}, which chooses randomly a cell
with the help of the method {\tt GenerCell2} and, next, 
the internal coordinates of the point within the cell
using {\tt MakeAlpha}.

The absolute coordinates of the MC event
are calculated and stored in the data member 
double-precision vector {\tt fMCvect}.
The MC weight is calculated using the procedure {\tt Eval}, which provides the
density distribution $\rho(x)$.

The MC event (double-precision vector) and its weight 
are available through getters {\tt GetMCvect} and {\tt GetMCwt}.

The user may alternatively call {\tt MCgenerate}, 
which invokes {\tt MakeEvent} and provides a MC event 
and its weight simultaneously.

\subsubsection{Finalize step methods}
The use of the method {\tt Finalize} is not mandatory.
It prints statistics and calculates the estimate of the
integral using the average weight from the MC run.
The amount of printed information depends on the values of {\tt fChat}.
For the normalization of the plots and integrals,
the user needs to know the exact value of $R'=\int \rho'(x) dx$, which is
provided by the method {\tt GetIntNorm} or {\tt Finalize}.

The actual value of the integrand from the MC series is provided
by {\tt GetIntegMC}.
Note that, for the convenience of the user,
{\tt GetIntNorm} provides $R'$ or an MC estimate of $R=\int \rho(x) dx$,
depending on whether the MC run was with variable weight
or weight $=1$ events.

Another useful finalization procedure
\begin{center}
\small
\begin{verbatim}
GetWtParams(const Double_t eps, Double_t &AveWt, Double_t &WtMax, Double_t &Sigma)
\end{verbatim}
\end{center}
\noindent
provides three parameters that characterize the MC weight distribution:
the average weight {\tt AveWt},
the ``intelligent'' maximum weight%
\footnote{The $\varepsilon$-dependent maximum weight is defined such
  that events with  $w>w^{\varepsilon}_{\max}$ contribute 
  an $\varepsilon$-fraction to the total integral.
  It is numerically more stable in the numerical evaluation
  than the one defined as the largest weight in the MC run.}
{\tt WtMax}~$=w^\varepsilon_{\max}$,
for a given value of {\tt eps}~$=\varepsilon$
and the variance {\tt sigma}~$=\sigma$.
In particular, in the case of $w=1$ events,
$w^\varepsilon_{\max}$ can be used as an input for the next MC run.

\subsubsection{Debug facility}
The {\tt TFoam} class includes method {\tt CheckAll} for the debugging purposes.
It checks the correctness of the pointers in the
doubly linked tree of cells (this can take time for large $N_c$).
Another debugging method {\tt PrintCells}
can be used at any stage of the calculation 
in order to print the list of all cells.

\begin{table}[!ht]
\centering
\begin{small}
\begin{tabular}{|l|p{11.0cm}|}
\hline
TFoamCell member & Short description  \\ 
\hline\hline
\hline
\multicolumn{ 2}{|c|}{ ``Static'' member, the same for all cells! }\\
\hline
  Short\_t   fDim     & Dimension of integration space\\
\hline
\multicolumn{ 2}{|c|}{ Linked tree organization}\\
\hline
  Int\_t     fSerial        & Serial number (index in fCells from TFoam class)\\
  Int\_t     fStatus        & Status (active or inactive)\\
  TRef    fParent        & Pointer to parent cell\\
  TRef    fDaught0       & Pointer to daughter 1\\
  TRef    fDaught1       & Pointer to daughter 2\\
\hline
\multicolumn{ 2}{|c|}{The best split geometry from the MC exploration}\\
\hline
  Double\_t fXdiv          & Factor $x$ of the cell split\\
  Int\_t    fBest          & The best edge candidate for the cell split\\
\hline
\multicolumn{ 2}{|c|}{Integrals of all kinds}\\
\hline
  Double\_t fVolume        & Cartesian volume of this cell\\
  Double\_t fIntegral      & Integral over cell (estimate from exploration)\\
  Double\_t fDrive         & Driver  integral $R_{\rm loss}$ for cell build-up\\
  Double\_t fPrimary       & Primary integral $R'$ for MC generation\\
\hline
\end{tabular}
\end{small}
\caption{\sf Data members of the {\tt TFoamCell} class.}
  \label{tab:TFCELLmembers}
\end{table}

\subsection{TFoamCell class}
The {\tt TFoamCell} class contains data and methods relevant to
a single cell object. Data members of the class are listed in
Table~\ref{tab:TFCELLmembers}. In comparison with {\tt FOAM} the number 
of data members is significantly reduced.
Most of the methods of the {\tt TFoamCell} class are setters and getters.
The non-trivial methods are {\tt GetHcub} and {\tt GetHSize},
which calculate the absolute position and size of hyperrectangles,
and {\tt CalcVolume}, which calculates the Cartesian volume of the cell.

The linked tree structure of {\tt TFoamCell} objects was not properly
treated by the ROOT automatic streamers, hence in the previous 
version of {\tt FOAM} the persistency has been achieved 
with the help of some workarounds --
namely pointers to cells in the linked list of cells
were replaced in {\tt FOAM} by the integer indexes%
\footnote{This workaround will be unnecessary
 after certain bugs have been corrected
 in the future implementation of the ROOT streamers.}.
In {\tt mFOAM} we go back to the
pointers, but instead of the raw C++ pointers we employ objects of
the special class of persistent pointers {\tt TRef} of ROOT.
This solution works very well, and as a consequence 
the method {\tt LinkCells}%
\footnote{{\tt LinkCells} and integer pointers in the {\tt TFoamCell} class
  were introduced in {\tt FOAM} as a ``workaround'' solution
  for certain problems with persistency of pointers in ROOT.
  It is still implemented in {\tt FOAM}
  as a void function in for the purpose of the backward compatibility
  in the user applications.}
from {\tt TFOAM} class became obsolete.
However, in the present implementation the memory consumption is
increased with respect to indexing using integers;
one cell now occupies 116 bytes of memory,
simply because objects of the {\tt TRef} class are composite objects.

\subsection{TRandom -- ROOT's collection of random-number generators}
\label{TRandom}

The full version 2.05 of {\tt FOAM}  uses its own internal random-number 
generator  called \mbox{RANMAR~\cite{Marsaglia:1990ig}}.
In {\tt mFOAM} it is replaced by the {\tt TRandom} class
interfacing to ROOT's internal library of the three random-number generators.
Two of them are rather simple generators, and we do not recommend their use in 
any serious applications.
We recommend to use its Mersenne Twister generator {\tt TRandom3},
which has huge period $2^{19937}-1$ and generally very good quality \cite{mtwistor}.
At the present moment the {\tt TRandom} package does 
not include any random-number  generator with the perfect (controllable) 
``randomness'',  such as RANLUX \cite{Luscher:1993dy} \cite{James:1993vv},
which is necessary  for certain applications%
\footnote{However,
  the authors of ROOT are planning to include RANLUX in the near future.}. 
Generally, we have decided to use {\tt TRandom},
because it meets our set of the minimal 
requirements for the library of random-number generators,
which can be characterized as follows:
\begin{itemize}
    \item Possibility to set (and reset) initial ``seed'' in the form of 
          just one integer.
    \item Availability of a method generating single uniform random number.
    \item Presence of a method generating series of uniform random numbers 
          in a single call.
    \item Possibility to record (disk-write) the complete status of 
          the random-number generator and restart it using this record.
          (This, of course, is assured by the persistency mechanism of the ROOT.)
  \end{itemize}
An advanced user of ROOT can also easily add his favourite
random-number generator with the same standardized  interface
(using inheritance from {\tt TRandom}).

The use of {\tt TRandom} is rather simple.
As an example lets us show the following line of code:
\begin{verbatim}
TRandom *PseRan   = new TRandom3(4357);  // Create random number generator
\end{verbatim}
which creates an instance of Mersenne Twister generator with the $seed=4357$.
Note that the {\tt TRandom} class includes many ``utility methods'',
however, only a small subset of them are used in {\tt mFOAM}.
For the detailed description of the {\tt TRandom}, class we 
refer the interested reader to the online ROOT documentation.

How to use a single {\tt TRandom} object for serving several 
objects of the {\tt TFoam} class is described in Section \ref{sec:extern}.

\section{Configuring {\tt mFOAM}}
\label{configuring}

\begin{table}[ht!]
\centering
\begin{small}
\begin{tabular}{|l|l|p{12.0cm}|}
\hline
Param. & Value & Meaning  \\ 
\hline\hline
  kDim     & 0$^*$    & Dimension of the integration space\\
  nCells   & 1000$^*$ & Maximum number of cells\\
  nSampl   & 200$^*$  & No. of MC events in the cell MC exploration\\
  nBin     & 8$^*$    & No. of bins in edge-histogram in cell exploration\\
  OptRej   & 1$^*$    & OptRej = 0, weighted; =1, $w=1$ MC events\\
  OptDrive & 2$^*$    & Maximum weight reduction\\
           & 1        & or variance reduction\\
  EvPerBin & 25$^*$   & Maximum number of effective $w=1$ events/bin\\
           & 0        & or counting of number of effective events/bin is inactive\\
  Chat     & 1$^*$    & =0,1,2 is the ``chat level'' in the standard output\\
  MaxWtRej & 1.1$^*$  & Maximum weight used to get $w=1$ MC events\\
\hline
\end{tabular}
\end{small}
\caption{\sf Nine principal configuration parameters and switches of the 
{\tt mFOAM} program. The default values are marked with the
 superscript star.}
  \label{tab:TFCELLparams}
\end{table}

At present {\tt mFOAM} has {\em nine principal configuration parameters}.
In addition, the user may optionally (re)define certain internal
configuration parameters of {\tt mFOAM} in order to inhibit and/or predefine the 
division geometry in the cell split. All of the nine principal 
parameters are listed in Table~\ref{tab:TFCELLparams}. 
They control all essential features of the program
and are preset to some meaningful default values, appropriate 
for the generation  of unweighted events.
The new inexperienced user of {\tt mFOAM} usually 
does not need to reset them. The only exception is the dimension
of integration space {\tt kDim}.
It is mandatory to set {\tt kDim} to a non-zero integer value
before invoking {\tt Initialize}.

In comparison with {\tt FOAM-2.05} two steering parameters were 
completely removed: {\tt nDim},  {\tt OptOrd}, as relevant only for simplical cells. 
The other three are hidden from the users eyes,
because their usefulness is rather limited.
Functionality of the program was frozen
for the following choice: {\tt OptPeek=0}, {\tt OptEdge=0 } and
{\tt OptMCell=1}. 
Finally, the default value of another optional input parameter
{\tt OptRej} switch is now set to 1 (weight $=1$ events), instead of 0.

If the user wants to redefine configuration parameters according
to his needs, then the relevant piece of code will look as follows:
\begin{verbatim}
   FoamX->SetkDim(     kDim);
   FoamX->SetnCells(   nCells);
   FoamX->SetnSampl(   nSampl);
   FoamX->SetnBin(     nBin);
   FoamX->SetOptRej(   OptRej);
   FoamX->SetOptDrive( OptDrive);
   FoamX->SetEvPerBin( EvPerBin);
   FoamX->SetMaxWtRej( MaxWtRej);
   FoamX->SetChat(     Chat);
\end{verbatim}

The user of {\tt mFOAM} can decide to inhibit the division in some
variables. This can be done with the method 
{\tt SetInhiDiv(Int\_t iDim, Int\_t InhiDiv)} of the class {\tt TFoam},
where {\tt iDim} is the index of the variable for which the inhibition is done
and {\tt InhiDiv} is the inhibition switch. This method should be used
before invoking {\tt Initialize}, after setting {\tt kDim}.
The relevant code may look as follows:
\begin{verbatim}
   FoamX->SetInhiDiv(0, 1); //Inhibit division of x_1
   FoamX->SetInhiDiv(1, 1); //Inhibit division of x_2
\end{verbatim}
The allowed values are {\tt InhiDiv=0,1} and
the default value is {\tt InhiDiv=0}.
Note that the numbering of integration variables
with the index {\tt iDim} starts from zero.
The inhibited variables are generated uniformly.

The user may also predefine divisions of the root cell
in certain variables using the method
{\tt SetXdivPRD(Int\_t iDim, Int\_t len, Double\_t xDiv[])}.
The relevant piece of the user code may look as follows:
\begin{verbatim}
   Double_t xDiv[3];
   xDiv[0]=0.30; xDiv[1]=0.40; xDiv[2]=0.65;
   FoamX->SetXdivPRD(0, 3, xDiv);
\end{verbatim}
Again, this should be done before invoking {\tt Initialize},
after setting {\tt kDim}. 

\section{Usage of the mFOAM package}
\label{sec:usage}

To begin work with the {\tt mFOAM} package, a user should have basic
knowledge of ROOT and the CINT interpreter.
Very good documentation of ROOT is available.
At this moment {\tt mFOAM} is already included 
in the ROOT standard distribution (beginning from version 4.04).
The ROOT package can be obtained from ROOT's web page%
\footnote{See http://root.cern.ch for more information.}. 
Precompiled binaries are also available as tar
archive files for many major platforms: PC computers with 
both Linux and MS Windows systems and workstations under UNIX.
All supported operating systems can be found on ROOT's home page. 
The installation process is straightforward and on most UNIX-like
systems amounts to unpacking the tarball file and setting 
environment variables: {\tt ROOTSYS}, which should point to the ROOT 
main directory and {\tt LD\_LIBRARY\_PATH} locating ROOT 
libraries. 

We strongly recommend to use binaries, which exactly match 
the user  operating system. If precompiled binaries for user 
system are not available, then a direct installation from source 
code is necessary. Source code can be obtained as a tarball 
or through the CVS repository. A detailed description of the 
configuration and compilation of the ROOT package is beyond 
the scope of this article. Therefore we refer the interested
user to ROOT's online documentation.

After successful installation, the shared library {\tt libFoam.so}
is present in the \${\tt ROOTSYS/lib} directory. This library can 
be loaded directly to ROOT by issuing the following command 
from CINT command line%
\footnote {Explicit loading of the {\tt mFOAM} 
  library is really needed in rare cases, when valid {\tt system.rootmap}
  file was not created after the compilation of source code with 
  the help of {\tt make map} command.}:
\begin{flushleft}
\hspace{0.8cm} \tt  root [0] .L \$ROOTSYS/lib/libFoam.so 
\end{flushleft}
From now on, the user will get an access to all {\tt mFOAM} classes
while interpreting/executing C++ scripts/programs under
the {\tt CINT} interpreter of {\tt ROOT},
or simply working interactively from the command line.

\subsection{Demonstration programs}
\label{example}

The user application program can be compiled/run
using one of the following three methods:
\begin{enumerate}
\item The user program is interpreted by {\tt CINT} of {\tt ROOT}.
  This simple method might be too slow in execution and will
  inhibit  the use of the persistency of the {\tt mFOAM} class.
\item The user program is compiled/linked in flight employing
  the Automatic Compiler of Libraries (ACLiC) facility of {\tt CINT}.
  This automatizes the process of compilation and linking
  and the persistency of the {\tt mFOAM} class is available.
  It is the preferred mode of work for medium and small-size applications.
\item Standard compile-link-run method.
  This method is well suited for large MC projects,
  which are run in the batch mode.
\end{enumerate}

We tried to provide the user with examples of all possible
compile/run methods. Demonstration scripts in the {\tt \$ROOTSYS/tutorials} 
directory cover the first two methods and show the basic features of {\tt mFOAM}. 
In addition there is a collection of simple programs showing how to 
build and run stand-alone applications.  They are distributed as a
{\tt mFoam-examples-1.2.tar}
file which is available from the authors web page.

\subsubsection{Examples in \$ROOTSYS/tutorials directory}
\label{tutorial}

Let us now describe in more detail some demonstration scripts in  
the {\tt tutorials} subdirectory of the ROOT distribution directory.
There are 3 demonstration programs there.
 
The first of them, {\tt foam\_demo.C}, demonstrates the full power of the 
{\tt mFOAM} compiled by ACLiC facility (scenario 2 above), 
showing all essential phases of its usage: initialization, the setting 
up of random-number generator a the distribution to be generated/integrated.
The examples of setting up optional input parameters are also shown.
Finally, MC generation and getting the value of the integral and other parameters
after MC generation are also demonstrated. 
This example is a slightly modified version of the analogous 
program in the {\tt FOAM} distribution~\cite{Jadach:2002kn}.
Let us explain the content of the {\tt foam\_demo.C} script.
After collection of headers we see the definition of the 
distribution to be generated/integrated:
\begin{verbatim}
   class TFDISTR: public TFoamIntegrand {
   public:
      TFDISTR();
      Double_t Density(Int_t, Double_t *){
      ......................
      }
      ClassDef (TFDISTR,1) //Class of testing functions for FOAM
   };
   ClassImp(TFDISTR)
   .....................
   TFoamIntegrand   *rho= new TFDISTR();
   FoamX->SetRho(rho);
\end{verbatim}
Class {\tt TFDISTR} inherits from the abstract class 
{\tt TFoamIntegrand}. 
Note the presence of the {\tt ClassImp} and {\tt ClassDef} statements, 
which tell ROOT to create an automatic streamer for this class.

The subsequent piece of the code creates the objects of the 
random-number generator, the integrand distribution and the 
{\tt mFOAM} object itself:
\begin{verbatim}
  TRandom    *PseRan     = new TRandom3();  // Create random number generator
  PseRan->SetSeed(4357);                    // Set seed
  TFoamIntegrand *rho= new TFDISTR();       // Create integrand distribution
  TFoam   *FoamX     = new TFoam("FoamX");  // Create MC simulator/generator
\end{verbatim}
Next, some configuration parameters of the {\tt TFoam} object
{\tt FoamX} are redefined before it is initialized (exploration):
\begin{verbatim}
  FoamX->SetkDim(kDim);        // mandatory!
  FoamX->SetnCells(nCells);    // optional
  FoamX->SetRho(rho);          // mandatory!
  FoamX->SetPseRan(PseRan)     // mandatory!		
  FoamX->Initialize();         // Initialize MC simulator/generator
\end{verbatim}
At this point the attention should be payed to the fact that just 
{\em after the exploration phase} 
the object of the {\tt mFOAM} class is written to file {\tt rdemo.root}:
\begin{verbatim}
    TFile RootFile("rdemo.root","RECREATE","histograms");
    .............
    FoamX->Write("FoamX");    // Writing mFOAM object on the disk
    .............
    RootFile.Write();
    RootFile.Close();
\end{verbatim}
Finally, the series of MC events are generated:
\begin{verbatim}
  for(loop=0; loop<NevTot; loop++)
  {
    FoamX->MakeEvent();           // generate MC event
    FoamX->GetMCvect( MCvect);    // get MC point
    MCwt=FoamX->GetMCwt();        // get MC weight
    ..........
  }
\end{verbatim}
The code ends up with the printouts of the value of the integral over PDF and some
other statistics concerning the MC run.
The user is invited to manipulate the configuration parameters
of {\tt mFOAM}. 
In particular we recommend to switch to weighted events ({\tt OptRej=0})
and change the number of cells {\tt nCells} in the initialization.

The {\tt foam\_demo.C} program is compiled, linked and executed
from the CINT shell by issuing the following commands:
\begin{flushleft}
 \hspace{0.8cm}   \tt \$ root  \\ 
 \hspace{0.8cm}   root [0] .L ../lib/libFoam.so \\
 \hspace{0.8cm}   root [1] .x foam\_demo.C+ \\
\end{flushleft}
Note that the suffix ``+'' instructs CINT to use the Automatic Compiler 
of Libraries (ACLiC) facility. In such a case the process of compilation
and linking is completely automatized. During the compilation 
phase the shared library  {\tt foam\_demo\_C.so} is  created, which 
contains the definition of the {\tt TFDISTR} class, together with 
its automatic streamers. This is exactly what we need for testing 
persistency. In the stand-alone application the class of the 
PDF would have to be directly compiled 
and put in the shared library for further use. Here it is done in 
a simplified way. 

The second small program, {\tt foam\_demopers.C}, demonstrates
the use of the persistency of the {\tt mFOAM} class.
It reads the {\tt mFOAM} object from the disk,
checks its consistency, prints out geometry of cells 
and starts generation of events. 
It can be interpreted directly by {\tt CINT}:
\begin{flushleft}
\hspace{0.8cm}  \tt \$ root  \\ 
\hspace{0.8cm}  \tt root [0] .x foam\_demopers.C\\
\end{flushleft}
The {\tt demo\_C.so} library, defining the {\tt TFDISTR} class,
is loaded at the run-time with the help of
\begin{verbatim}
   gROOT->ProcessLine(".L foam_demo.C+")
\end{verbatim}
in the code. 
The user may verify that the output from it is {\em exactly the same} 
as the analogous output of  {\tt foam\_demo.C}. 
This illustrates the fact that the 
{\tt mFOAM} object, the MC simulator, can be dumped into the disk at any moment and 
it resumes its functioning after reloading it from the disk, as if there was 
no disk-write and disk-read at all.

The other macro {\tt foam\_kanwa.C} is a simplified
shorter version of {\tt foam\_demo.C}, without any unnecessary 
modification of the configuration parameters of the {\tt mFOAM}
(they are internally set to sensible default values).
This macro might be useful for the first-time user of the {\tt mFOAM}.
On the other hand, this program adds a simple example of the graphics
using {\tt ROOT}; the 2-dimensional distribution of the
produced MC events is shown dynamically on the screen,
as the accumulated MC statistics grows.
Notice the use of the {\tt TApplication} object, in order to stabilize
the picture on the screen in the execution.
This macro can be executed/interpreted (scenario 1)
directly by means of typing:
\begin{flushleft}
\hspace{0.8cm}  \tt  \$ root \\ 
\hspace{0.8cm}  \tt  root [0] .x foam\_kanwa.C \\
\end{flushleft}
The example output from running {\tt foam\_kanwa.C} is reproduced 
in the appendix. Simulation will start and then
a plot of the distribution function will pop-up on the graphical 
{\em canvas} on the screen. 
The execution is noticeably slower, as is always the case
for the interpreted programs.
The main difference with the {\tt foam\_kanwa.C} is in the
distribution function, which is now defined simply as a global
function {\tt Camel2}. 
It is made accessible to the {\tt mFOAM} object {\tt FoamX}
in the following line of code: 
\begin{verbatim}
    FoamX->SetRhoInt(Camel2);  
\end{verbatim}
Another difference is that the shared library of {\tt mFOAM}
is loaded with the following explicit instruction:
\begin{verbatim}
    gSystem->Load("libFoam.so");
\end{verbatim}
instead of the linking procedure. This instruction is not really needed
if ROOT is already aware of the location of the {\tt mFOAM} library. 

In some of the above examples we could not exploit the persistency of the 
ROOT objects.
This is because of the restrictions in CINT, which does not
allow an interpreted function to inherit from the {\tt TObject} class.
This is the reason why, in these examples where PDF is the
global function, the automatic streamer cannot be generated.
Even if one would write the {\tt mFOAM} object 
on the disk, the information  about the PDF will be lost. 
Of  course, the user may always go back
to one of the compilation methods and enjoy full persistency of the {\tt mFOAM} 
objects. In addition to better persistency, the compiled applications 
have the advantage of being significantly faster in the execution.

\subsubsection{More advanced examples of the use of {\tt mFOAM}}
\label{examps}
Let us now describe in more detail some examples of the use of {\tt mFOAM} 
classes in stand-alone applications (scenario 3).
It may be of interest to more advanced users,
who plan to use {\tt mFOAM} as part of their large-scale Monte Carlo projects.
It is assumed that {\tt ROOT} is installed and the 
environment  variable {\tt ROOTSYS} is properly set.

After unpacking the distribution file {\tt mFoam-examples-1.2.tar}
one should execute the {\tt configure} script:
\begin{flushleft}
 \hspace{0.8cm}  \tt \$ cd mFoam-examples-1.2 \\ 
 \hspace{0.8cm}  \tt \$ ./configure \\
\end{flushleft}
which inspects the system  configuration and looks up for the {\tt ROOT}
library, then generates the {\tt Makefile} file.
Version 4.04 of {\tt ROOT} or later is required.
The {\tt configure} script can fail for
many reasons. In that case the user should first check if the {\tt ROOTSYS} 
environment variable indeed points to {\tt ROOT} installation location. 
The default behaviour of the {\tt configure} script can be changed by additional 
command line parameters and environment variables. It may be useful if 
the computer is equipped with a compiler other than {\tt gcc}. A full list of 
available options is displayed by the {\tt 'configure -h'} command.

The {\tt configure} script and the accompanying configuration files
were generated%
\footnote{The distribution directory with the {\tt configure} script
  was created with  the command sequence
  {\tt (autoreconf -i; ./configure; make dist)},
  activating the directive {\tt AM\_MAINTAINER\_MODE} in {\tt config.in}.
}
using {\tt automake} tools version 1.91. In case the user wants to 
re-create the {\tt configure} script and the accompanying files, version 1.61 
or later of {\tt automake} is needed.

To compile and link these codes one should type the following:
\begin{flushleft}
 \hspace{0.8cm}  \tt \$ make \\ 
 \hspace{0.8cm}  \tt \$ make install \\
\end{flushleft}
We have successfully tested  the installation of {\tt mFOAM-examples} 
on computers  with  several variants of the Linux operating system: 
CERN Scientific Linux SLC3, Red Hat Linux 7.3, Fedora Linux FC3. The code is 
highly portable and  we think that it should compile without any problems 
on all other systems supported by ROOT developers. In rare case, 
certain minor modifications of the source code may be necessary.

After successful compilation one can run demonstration programs
with the following commands:
\begin{verbatim}
    make kanwa
    make demo
    make testpers
\end{verbatim}

The content and functionality of the programs {\tt demo.cxx} 
and {\tt kanwa.cxx} are the same as those of their macro counterparts
{\tt foam\_demo.C} and {\tt foam\_kanwa.C} described above.
The code of these programs can serve as a useful template
for the user applications.

The command {\tt make testpers} runs an advanced test of persistency
with two generator objects served by one central random generator.
In this example two classes of MC event generators {\tt TGenMC} and {\tt TGenMC2}
are defined and the corresponding library {\tt libTGenMC.so} is created.
An object of each MC event generator uses one own object of {\tt mFOAM}
class and one external object of the class {\tt TRandom} --
the central RNG. 

In the program {\tt Main.cxx}, two objects of 
the class {\tt TGenMC} and {\tt TGenMC2} are created.
Also a single central RNG object is allocated
and made available to both MC generators.
All three objects are written into a disk file
and used to generate 200k MC events, using each of the two MC generators.

The other program {\tt MainW.cxx} reads all three objects from the disk file,
reassigns the central RNG to {\tt mFOAM} objects inside the two MC event generators;
again, 200k MC events are generated, using each of the two MC generators.
Since the disk-write in {\tt Main} was done after initialization and
before MC generation, the MC series of the events from {\tt MainW}
should be the same as from {\tt Main}. 
This is checked by ``diffing'' two files which record first 15 events from
{\tt Main} and {\tt MainW} correspondingly.
We find that their content is identical, and this 
provides an empirical proof that this complicated setup
of the two MC event generators using two {\tt mFOAM} objects and
single central RNG is surviving the disk-write and disk-read operations
without any loss of its functionality.

The compile--link--execute chain for the tandem of {\tt Main} and {\tt MainW}
programs and ``diffing'' output files is realized by the single
command `{\tt make testpers}'.

The above example of organization with the single central RNG
is well suited for any large Monte Carlo projects
with many {\tt mFOAM} objects and many Monte Carlo sub-generators served
by the single central RNG.

The other interesting feature of the above examples is the implementation
of the PDF as the {\tt Density} method of the {\tt TGenMC2} class.
In our example the {\tt TGenMC2} class inherits from {\tt TFoamIntegrand}.
Consequently, the {\tt Density} function is provided to the {\tt mFOAM} object
(which is the member of the {\tt TGenMC2} class) as {\tt this}.
In the other MC generator of the class {\tt TGenMC}, the PDF is defined
in an object of the separate class {\tt TFDISTR} and the PDF of this class
is allocated and its pointer is assigned to the {\tt mFOAM} object
in the {\tt TGenMC} object during its initialization.

The above test demonstrates a few fairly complicated examples
of how to organize the relation between several {\tt mFOAM} objects, RNGs and PDFs
within the MC project.
However, it does not cover all possible situations.
In the next section we shall discuss this issue in a general case and we
shall argue that object of the {\tt mFOAM} class are able to cope with all possible
scenarios in an efficient and transparent way.

\subsection{External RNG and PDF objects and\\ the implementation of persistency}
\label{sec:extern}
Persistency is undoubtedly a very valuable feature of the objects of 
the class {\tt TFoam}, and of ROOT objects in general.
It is therefore worthwhile to clarify certain features of its
implementation, which the user should know and consider before
attempting to exploit the persistency
of {\tt mFOAM} objects in any advanced/sophisticated applications.

As we have seen in the explicit examples of the previous section,
the critical issue in this context is the treatment of the two
external objects, which every given object of the class {\tt TFoam} needs
in order to function properly: the object of random number generator (RNG)
and the object providing the probability distribution function (PDF).
These two objects have to be provided to the object of the class {\tt TFoam}.
In the previous section, we have shown
most typical case, when one deals with only one instance (object)
of the above classes -- this was quite straightforward to organize.

In more advanced applications we have to be prepared to deal
with the situations in which we deal with many (hundreds)
of objects of the class {\tt TFoam},
all of them using single {\em central} RNG object
(or a few of them) and possibly generating many different PDFs.
In this case if one wants to profit fully from the persistency,
such a complicated set of interrelated objects
of the three types has to emerge fully operational after the disk-write
and disk-read operations.
This turns out to be a nontrivial task to realize in practice.

We claim that the way we interface an object of the class {\tt TFoam}
with the two ``satellite'' RNG and PDF objects of the {\tt TRandom}
and {\tt TFoamIntegrand} classes allows us to deal
with any arbitrarily complicated set of interrelated object,  while
correctly implementing the persistency in {\em all} such situations.

First of all, the two external objects, RNG and PDF, are external
in this sense that the {\tt new} operator allocating them is placed outside the
{\tt TFoam} code and the object of the class {\tt TFoam}
knows only their pointers.
Hence, the important question related to the persistency implementation
using ROOT (the problem is however more general) can be 
immediately formulated:
Whose responsibility is to {\em re-create} these two objects 
in the process of the disk-read?
First possible solution is that this task is handled by the automatic streamer
of the object of the class {\tt TFoam}, 
which would re-create the objects RNG and PDF%
\footnote{With the help of their own streamers.}.
Their actual pointer should then be exported to
any other objects, which need legitimately an access to them.
This second possibility is to inhibit the re-creation of the objects RNG and PDF
by the streamers of the class {\tt TFoam}.
ROOT allows this to be done%
\footnote{It is done by means of adding the comment 
  {\tt $\backslash\backslash$!} at the end of the line in which the pointer
  to an object is defined.}. 
In the latter case it would be the sole responsibility of the user to store the
two external objects  RNG and PDF
into disk separately, read them separately,
and provide their pointers to the object of the class {\tt TFoam}
after the disk-read operation.

The first option looks attractive because of its simplicity.
It is definitely an optimal one
in the most common case of just three objects -- hence we would like to 
implement this scenario as the basic one.
However, this solution will fail when several objects
of the {\tt TFoam} class are served by the single RNG object, a quite common case
in any bigger MC projects.
In this case, the disk-read operation (done by ROOT streamers)
will clone many independent identical RNG objects,
each one for every object of the class {\tt TFoam}.
This is clearly undesirable.

The situation with a set of several PDF objects serving one or more
{\tt TFoam} objects is even more subtle. 
On the one hand, one may argue that since the distribution
of a given PDF object is essentially memorized inside a given {\tt TFoam} object,
a genuine one-to-one association among them should be maintained.
Hence, the PDF object
should be ``owned'' by the {\tt TFoam} object, during the disk-write
and disk-read, as in the first scenario.
On the other hand, we shall sometimes 
deal with the situations with a single PDF object
serving several {\tt TFoam} objects; either because it needs huge memory,
or it is very slow in execution (its execution is a two-step process),
or it is not a genuine C++ object, but rather a ``wrapper''
to another non-OOP (Fortran) program.
In such a case it is better to handle PDF objects outside the {\tt TFoam} object,
as in the second scenario.
Summarizing, the treatment of the RNG and PDF objects should be quite similar,
and the possibility of keeping/controlling
both of them outside the {\tt TFoam} object should be optionally available.
In other words, 
we would ideally need both above solution for both RNG and PDF objects:
The first solutions, for simple applications 
and the second one for advanced applications.

The actual method of handling the RNG and PDF external objects in the {\tt TFoam}
class allows the user to implement both above scenarios.
It is done in the following way: the RNG and PDF objects are always 
created for the first time outside the {\tt TFoam} object, as already described.
Their pointers are transferred into the {\tt TFoam} object
as the arguments of {\tt Initialize(RNG,PDF)}.
Alternatively it is done with the help of the two 
dedicated setters {\tt SetPseRan(RNG)} and {\tt SetRho(PDF)},
before invoking {\tt Initialize()}.
At first sight, it seems that we follow the first solution, especially that
we do not inhibit the re-creation of the ``private copy'' of the objects RNG and PDF 
by the {\tt TFoam} object (by its streamer) during the disk-read.
Indeed the first solution is available in this way.
Restricting the discussion to RNG objects,
the second scenario can be implemented as follows:
first, disk-write and disk-read of the RNG object is done by the user,
then, after the disk-read of all {\tt TFoam} objects, the pointers of the RNG object
inside {\tt TFoam} objects are reassigned to the RNG object,
using a dedicated setter method, see also the examples of section \ref{examps}.
In order to avoid a memory leak,
the setter which is used to reassign the pointer to external RNG 
has to destroy the existing ``ghost'' RNG object,
which has been unnecessarily created during the disk-read operation 
of every {\tt TFoam} object.
The method {\tt ResetPseRan(RNG)} is introduced exactly for this purpose.
The analogous setter method of destroying the existing PDF object 
and reassigning its pointer is the method {\tt ResetRho(PDF)}
of the {\tt TFoam} class.
Obviously, the RNG and PDF objects are treated in the same way.

The above solution is efficient, transparent and useful in almost all cases.
It will not be satisfactory in the case where creating 
and destroying a PDF object takes 
an extremely long time and/or huge memory (no such problem with the RNG objects).
In such a case a simple modification of the source code
of the {\tt TFoam} class (inhibiting the storage of the PDF object)
will be a more economic solution; however, it requires recompiling 
the {\tt TFoam} library.

\section{Conclusions}

We present all users of the {\tt FOAM} package with its new version {\tt mFOAM}.
We have payed special attention to
making it more user-friendly, so that it provides with less effort
solutions of many every-day problems in the MC simulation.
This may, hopefully, attract new users, 
especially those who already use ROOT in their work.
We also hope for feedback from them, to be used
in the further improvements in the user interface to 
both {\tt FOAM} and {\tt mFOAM}.

\section*{Acknowledgements}

We are very grateful to R. Brun and F. Rademakers for their help
in achieving better integration of the {\tt mFOAM} code with ROOT,
and many related discussions.
We warmly acknowledge the help of Piotr Golonka in setting up the
example distribution directory.
We thank to ACK Cyfronet AGH Computer Center
for granting us access to their supercomputers and PC clusters
funded by computational grants:
MNiI/SGI2800/IFJ/009/2004,
MNiI/HP\_K460-XP/IFJ/009/2004,
EU project CrossGrid IST-2001-32243,
and KBN Grant SPUB-M 620/E-77/SPB/5PR UE/DZ 224/2002-2004,
which were helpful while testing {\tt mFOAM}.

\providecommand{\href}[2]{#2}

\appendix
\newpage
\section{APPENDIX}
The code of the example macro {\tt foam\_kanwa.C} looks like:

\vspace{-3mm}
{\footnotesize
\begin{verbatim}
//
// This program can be executed from the command line:
//     
//      root -l foam_kanwa.C
//_____________________________________________________________________________
Int_t kanwa(){
  cout<<"--- kanwa started ---"<<endl;
  gSystem->Load("libFoam.so");
  TH2D  *hst_xy = new TH2D("hst_xy" ,  "x-y plot", 50,0,1.0, 50,0,1.0);
  Double_t *MCvect =new Double_t[2]; // 2-dim vector generated in the MC run
  TRandom     *PseRan   = new TRandom3();  // Create random number generator
  PseRan->SetSeed(4357);                   // Set seed
  TFoam   *FoamX    = new TFoam("FoamX");   // Create Simulator
  FoamX->SetkDim(2);         // No. of dimensions, obligatory!
  FoamX->SetnCells(500);     // No. of cells, can be omitted, default=2000
  FoamX->SetRhoInt(Camel2);  // Set 2-dim distribution, included below
  FoamX->SetPseRan(PseRan);  // Set random number generator
  FoamX->Initialize();       // Initialize simulator, may take time...
  // From now on FoamX is ready to generate events according to Camel2(x,y)
  for(Long_t loop=0; loop<100000; loop++){
    FoamX->MakeEvent();            // generate MC event
    FoamX->GetMCvect( MCvect);     // get generated vector (x,y)
    Double_t x=MCvect[0];
    Double_t y=MCvect[1];
    if(loop<10) cout<<"(x,y) =  ( "<< x <<", "<< y <<" )"<<endl;
    hst_xy->Fill(x,y);
  }// loop
  Double_t MCresult, MCerror;
  FoamX->GetIntegMC( MCresult, MCerror);  // get MC integral, should be one
  cout << " MCresult= " << MCresult << " +- " << MCerror <<endl;
  // now hst_xy will is plotted, visualizing generated distribution
  TCanvas *cKanwa = new TCanvas("cKanwa","Canvas for plotting",600,600);
  cKanwa->cd();
  hst_xy->Draw("lego2");
  cout<<"--- kanwa ended ---"<<endl;
}//kanwa
//_____________________________________________________________________________
Double_t sqr(Double_t x){return x*x;};
Double_t Camel2(Int_t nDim, Double_t *Xarg){
// 2-dimensional distribution for Foam, normalized to one (within 1e-5)
  Double_t x=Xarg[0];
  Double_t y=Xarg[1];
  Double_t GamSq= sqr(0.100e0);
  Double_t Dist= 0;
  Dist +=exp(-(sqr(x-1./3) +sqr(y-1./3))/GamSq)/GamSq/TMath::Pi();
  Dist +=exp(-(sqr(x-2./3) +sqr(y-2./3))/GamSq)/GamSq/TMath::Pi();
  return 0.5*Dist;
}// Camel2
//_____________________________________________________________________________
                                                                                
\end{verbatim}
}


Macro {\tt foam\_kanwa.C} produces the following output:
{\footnotesize
\begin{verbatim}
--- kanwa started ---
FFFFFFFFFFFFFFFFFFFFFFFFFFFFFFFFFFFFFFFFFFFFFFFFFFFFFFFFFFFFFFFFFFFFFFFFFFFFFFFF
F                                                                              F
F                   ****************************************                   F
F                    ******      TFoam::Initialize    ******                   F
F                   ****************************************                   F
F                                                      FoamX                   F
F    Version =           1.02M   =                   Release date:  2005.04.10 F
F       kDim =          2 =  Dimension of the hyper-cubical space              F
F     nCells =        500 =  Requested number of Cells (half of them active)   F
F     nSampl =        200 =  No of MC events in exploration of a cell          F
F       nBin =          8 =  No of bins in histograms, MC exploration of cell  F
F   EvPerBin =         25 =  Maximum No effective_events/bin, MC exploration   F
F   OptDrive =          2 =  Type of Driver   =1,2 for Sigma,WtMax             F
F     OptRej =          1 =  MC rejection on/off for OptRej=0,1                F
F   MaxWtRej =             1.1   =       Maximum wt in rejection for wt=1 evts F
F                                                                              F
FFFFFFFFFFFFFFFFFFFFFFFFFFFFFFFFFFFFFFFFFFFFFFFFFFFFFFFFFFFFFFFFFFFFFFFFFFFFFFFF
2222222222222222222222222222222222222222222222222
FFFFFFFFFFFFFFFFFFFFFFFFFFFFFFFFFFFFFFFFFFFFFFFFFFFFFFFFFFFFFFFFFFFFFFFFFFFFFFFF
F                                                                              F
F                    ***  TFoam::Initialize FINISHED!!!  ***                   F
F     nCalls =      99800 =            Total number of function calls          F
F     XPrime =       1.3922344   =     Primary total integral                  F
F     XDiver =      0.39314276   =     Driver  total integral                  F
F   MCresult =      0.99909163   =     Estimate of the true MC Integral        F
F                                                                              F
FFFFFFFFFFFFFFFFFFFFFFFFFFFFFFFFFFFFFFFFFFFFFFFFFFFFFFFFFFFFFFFFFFFFFFFFFFFFFFFF
(x,y) =  ( 0.26506687, 0.37983892 )
(x,y) =  ( 0.65874831, 0.76719268 )
(x,y) =  ( 0.6405293, 0.73329734 )
(x,y) =  ( 0.29933616, 0.37537068 )
(x,y) =  ( 0.31228105, 0.39614503 )
(x,y) =  ( 0.71258758, 0.64969589 )
(x,y) =  ( 0.34830539, 0.38099167 )
(x,y) =  ( 0.26990382, 0.42078097 )
(x,y) =  ( 0.35661486, 0.41847364 )
(x,y) =  ( 0.5557375, 0.62757837 )
 MCresult= 1.0004446 +- 0.00080837425
--- kanwa ended ---
\end{verbatim}
}

\newpage
In addition the same script {\tt foam\_kanwa.C} produces the following 
plot (canvas), which pops up on the screen.
\vspace{10mm}

\begin{center}
\epsfig{file=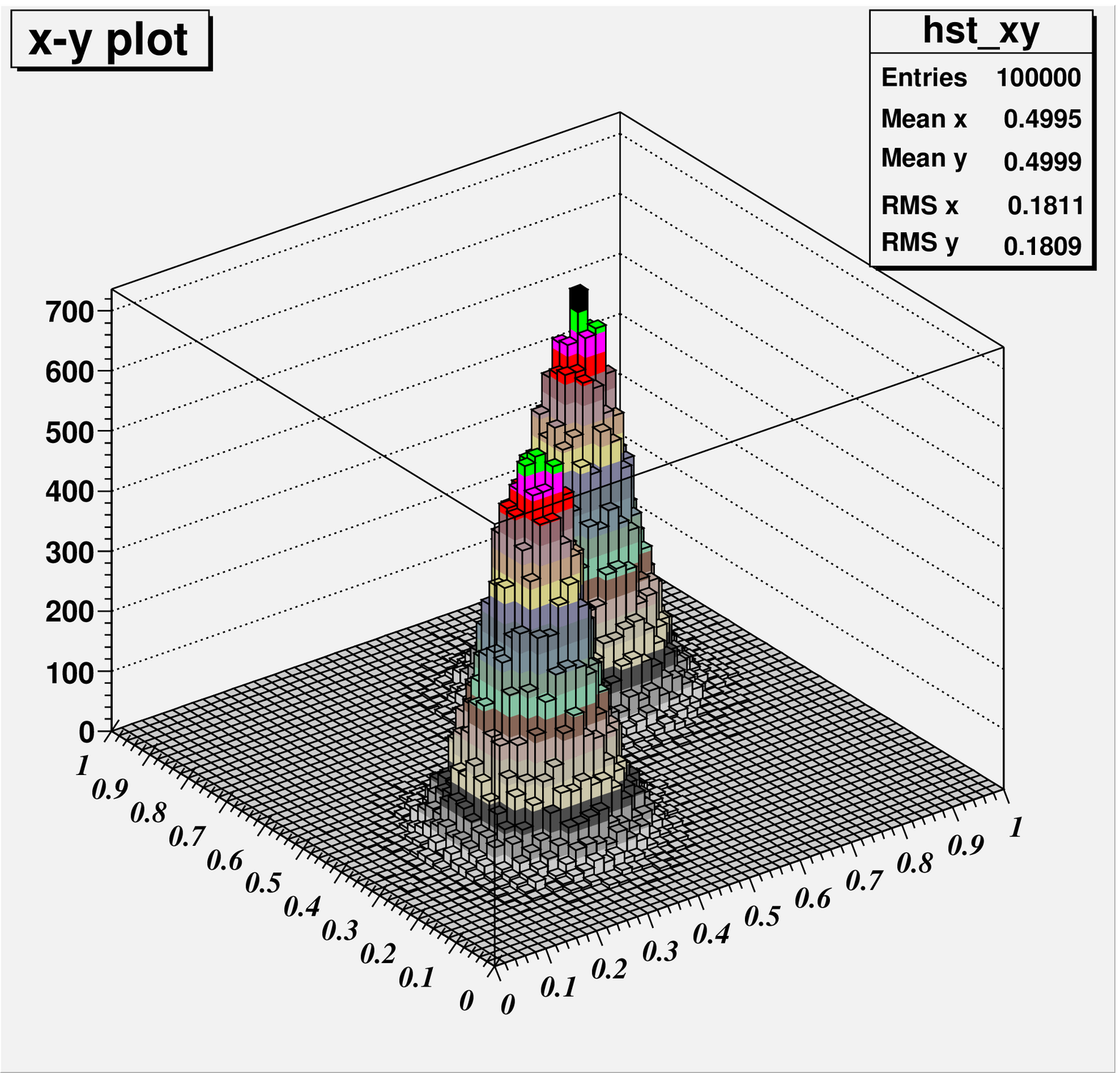,width=150mm}
\end{center}

\end{document}